\documentclass[11pt,aps,nofootinbib,preprint,superscriptaddress]{revtex4}
\usepackage{color}
\usepackage{graphicx}
\usepackage{amsmath,amssymb,amsthm,amscd}
\newcommand{\IP}{\relax{\rm I\kern-.18em P}}
\newcommand{\IR}{\relax{\rm I\kern-.18em R}}

\def\be{\begin{equation}}
\def\ee{\end{equation}}

\def\bea{\begin{eqnarray}}
\def\eea{\end{eqnarray}}

\def\nn{\nonumber}
\def\ov{\over}
 
\def \ddel#1#2{\frac{\partial^2 #1}{\partial #2^2}}

\def\vev#1{\langle #1 \rangle}

 \usepackage{setspace} 

\newcommand{\CO}{{\cal O}}

\def\IR{{\mathbb R}}

\def\IP{{\mathbb P}}

\newcommand{\tr}{{\rm Tr}}

\newcommand{\ba}{\begin{aligned}}
\newcommand{\ea}{\end{aligned}}

\newcommand{\ben}{\begin{eqnarray}\displaystyle}
\newcommand{\een}{\end{eqnarray}}

\begin{document}
\preprint{TIFR/TH/06-27}

\title{Gauge theory description of the fate of the small Schwarzschild blackhole}

\author{Spenta R. Wadia}
\email{wadia@theory.tifr.res.in}
\affiliation{Tata Institute of Fundamental Research, Homi Bhaba Road, Mumbai-400005, India}

\vspace*{5.0ex}

\begin{abstract}
In this talk we discuss  the fate of the small Schwarzschild blackhole of $AdS_5\times S^5$ using the AdS/CFT correspondence at finite temperature.
The third order $N = \infty$ phase transition in the gauge theory corresponds to the
blackhole-string transition. This singularity is resolved using a
double scaling limit in the transition region. 
The phase transition becomes a smooth crossover where multiply wound Polyakov lines condense. In particular the density of states is also smooth at the crossover. We discuss the implications of our results for the singularity of the Lorenztian section of the small Schwarzschild blackhole.
(\it {Talk given at the 12th Regional conference in Islamabad, Pakistan, based on hep-th/0605041})

\end{abstract}

\maketitle

\section {Blackholes and string theory}

Over the past several years the subject of blackhole physics has had a symbiotic relationship with string theory. Blackholes provide a perfect theoretical laboratory to develop and test string theory. In turn string theory provides a consistent framework to deduce blackhole thermodynamics from quantum statistical mechanics. For a review see \cite{David:2002wn}.

This relationship has been most fruitful in the case of the 5-dim. supersymmetric blackhole solution of Strominger and Vafa whose microscopic degrees of freedom are D1 and D5 branes. Here it was shown that the Bekenstein-Hawking formula for blackhole entropy gave the same result as Boltzmann's formula. The Strominger-Vafa blackhole is extremal and its microscopic degeneracy is that of the D1/D5 system in its ground state. Small excitations around the ground state correspond to small deviations from extremality and a small non-zero temperature. Such blackholes emit Hawking radiation which can be calculated as an emission/absorption process in the microscopic theory. All these calculations of the microscopic theory match with semi-classical general relativity because of the high degree of supersymmetry in the system. The effective field theory of the D1/D5 system is a symmetric product 2-dim. superconformal field theory with (4,4) supersymmetry. Its central charge determines the thermodynamics of the blackhole and its 2-point functions compute the Hawking radiation formulas. The key reason why the microscopic calculations remain applicable in the gravity domain is because the high degree of supersymmetry circumvents the strong coupling problem. 

Our aim in this talk is to present some progress in trying to understand blackholes which are far from being supersymmetric, using the AdS/CFT correspondence \cite{Alvarez-Gaume:2006jg}. Here unlike the supersymmetric case we have to face the strong coupling and intermediate coupling problems of the non-abelian gauge theory. There has been a lot of work in connection with blackholes in $AdS_5$. Almost all this work is qualitative and deals in arriving at a phase diagram for the gauge and bulk theory in the large N limit using both weak coupling expansions in the gauge theory and the supergravity correspondence at strong coupling \cite{Witten:1998zw}, \cite{Sundborg:1999ue}, \cite{Polyakov:2001af}, \cite{Aharony:2005bq}, \cite{Alvarez-Gaume:2005fv}, \cite{Basu:2005pj}.

\section{Small 10-dim. Schwarzschild blackhole in $AdS_5\times S^5$}

The problem of the fate of small Schwarzchild blackholes is important to understand, in a quantum theory of gravity. In a 
unitary theory this problem is the same as the formation of a small blackhole. An understanding of this phenomenon 
has bearing on the problem of spacelike singularities in quantum gravity and also (to some extent) on 
the information puzzle in blackhole physics. It would also teach us something about non-perturbative string physics.

We focus attention on the fate of small 10-dim. Schwarzschild blackhole in $AdS_5\times S^5$, which (for reasons we will explain) is amenable to a precise quantitative treatment under reasonable assumptions \cite{Alvarez-Gaume:2006jg}. In order to avoid confusion we mention that the blackhole we are referring to, is the small Schwarzchild blackhole of 10-dim. spacetime rather than the small blackhole in $AdS_5$. The latter, unlike our case, is uniformly spread over $S^5$, and hence has the Gregory-Laflamme instability (\cite{Hubeny:2002xn}).

In the past Susskind \cite{Susskind:1993ws}, Horowitz and Polchinski (SHP) \cite{Horowitz:1996nw} and others 
\cite{Sen:1995in,Bowick:1986km,Bowick:1985af} have discussed this, in the framework of string theory, as a blackhole-string 
transition or more appropriately a crossover. Their proposal was that this crossover is parametrically smooth and it simply 
amounts to a change of description of the same quantum state in terms of degrees of freedom appropriate to the strength 
of the string coupling. The entropy and mass of the state change at most by $o(1)$. By matching the entropy formulas for 
blackholes and perturbative string states, they arrived at a crude estimate of the small but non-zero string coupling at the crossover. 

The SHP description is difficult to make more precise because 
the blackhole-string crossover occurs in a regime where the curvature of the blackhole 
is $o(1)$ in string units, so as to render the supergravity description invalid. It is also clear that besides $l_s$ related effects, the string coupling is non-zero and its effects have to be taken into account. 
Presently our understanding of string theory is not good enough for us to make a precise and quantitative discussion of the crossover. Hence we will discuss the problem using the AdS/CFT 
correspondence. 

When the horizon of this blackhole approaches the string scale $l_s$, we expect the supergravity (geometric) 
description to break down and be replaced by a description in terms of degrees of freedom more appropriate at 
this scale. Presently we have no idea how to discuss this crossover in the bulk IIB string theory. 
Hence we will discuss this transition and its smoothening in the framework of a general finite temperature 
effective action of the dual $SU(N)$ gauge theory on $S^3\times S^1$. In fact it is fair to say that in the crossover region we are really using the gauge theory as a definition of the non-perturbative string theory.

The use of the AdS/CFT correspondence for studying the blackhole-string crossover requires that there is a 
description of small Schwarzschild blackholes as solutions of type IIB string theory in $AdS_5 \times S^5$. 
Fortunately, Horowitz and Hubeny \cite{Horowitz:2000kx} have studied this problem with a positive conclusion. 
This result enables us to use the boundary gauge theory to address the crossover of the small Schwarzschild blackhole into a state described in terms of 'stringy' degrees of freedom. 
\par
\section{Gauge theory and effective action }

At finite temperature we are dealing with the $\cal N$=4, $SU(N)$ SYM theory on $S^3\times S^1$. At large but finite $N$, since $S^3$ is compact, the partition function and all correlation functions are 
smooth functions of the temperature and other chemical potentials. There is no phase transition. 
However in order to make a connection with a dual theory of gravity, which has infinite number of degrees 
of freedom, we have to take the $N\rightarrow \infty$ limit and study the saddle point expansion in 
powers of $\frac{1}{N}$. It is this procedure that leads to non-analytic behavior. It turns out that by taking 
into account exact results in the $\frac{1}{N}$ expansion it is possible to resolve this singularity and recover a smooth crossover in a suitable double scaling limit. 

The gauge theory is very hard to deal with as we have to solve it in the $\frac{1}{N}$ expansion for large but finite values of the 'tHooft coupling $\lambda$. Inspite of this there is a window of opportunity to do some precise calculations because it can be shown that the 
effective action of the gauge theory at finite temperature can be expressed entirely in terms of the Polyakov loop 
which does not depend on points on $S^3$: $U = P \exp \Bigl(i \int_0^{ \beta} A_{0} d \tau \Bigr)$, where $A_{0} (\tau) $ is the zero mode of the time component of the gauge field on $S^3$. This is a single $N\times N$ unitary matrix, albeit with a complicated interaction among the winding modes $\tr U^n$. 

Two comments are in order:

i) Since the 10-dimensional blackhole sits at a point in $S^5$, one may be concerned about the spontaneous breaking of 
$SO(6)$ R-symmetry and corresponding Nambu-Goldstone modes. However using a supergravity analysis (\cite{Alvarez-Gaume:2006jg}), we have concluded that the symmetry is not spontaneously broken. Instead we have to introduce collective coordinates for treating the zero modes associated with this symmetry. 

ii) The circumstance, that the order parameter $U$ in the gauge theory is a constant on $S^3$, matches well on the supergravity side with the fact that all the zero angular momentum 
blackhole solutions are also invariant under the $SO(4)$ symmetry of $S^{3}$. The blackhole may be localized in $S^5$, 
but it does not depend on the co-ordinates of $S^3$ on which it is uniformly spread. The coefficients of the effective action depend upon the temperature, the `t Hooft coupling $\lambda$ and the vevs of the scalar fields.

\section{Effective action and multi-trace unitary matrix model}

The partition function of the gauge theory can now be written as a general unitary matrix model,
\bea
\label{genz}
  Z (\lambda,T)= \int dU \, e^{ S (U)}
\eea
Gauge invariance requires that the effective action of $U$ be expressed in terms of products of $\tr U^n$, with $n$ an integer, since these are the only gauge invariant quantities that can be constructed from $A_{0}$ alone. $S(U)$ also has a $Z_N$ symmetry under $U \to e^{{2 \pi i \ov N}} U$

These requirements fix the effective action to be of the form
\bea
S (U,U^{\dagger}) = \sum_{i=1}^p a_i \tr \, U^i  \tr \, U^{\dagger i} + \sum_{\vec k, \vec k'} \alpha_{\vec k , \vec k'} \Upsilon_{\vec k} (U) \Upsilon_{\vec k'}(U^{\dagger}),
\label{genaction}
\eea
where $\vec k$, $\vec k'$ are arbitrary vectors of nonnegative entries, and
\be
\label{upsop}
\Upsilon_{\vec k} (U)=\prod_j \Bigl( \tr \, U^j \Bigr)^{k_j}.
\ee
$a_i$ and $\alpha_{\vec k , \vec k'}$ are parameters that depend on the temperature $\beta^{-1}$ and the coupling $\lambda$.
Reality of the action (\ref{genaction}) requires $\alpha_{\vec k \vec k'} =\alpha^*_{\vec k' \vec k}$. In fact, using the explicit perturbative rules to compute $S(U,U^{\dagger})$ in (\ref{genaction}), one can show that the $\alpha_{\vec k \vec k'}$ are real, therefore 
\be
\label{alphacond}
\alpha_{\vec k \vec k'} =\alpha_{\vec k' \vec k}.
\ee

\par

\section{A lemma in matrix theory}

The general unitary matrix model can be analyzed due a lemma that enables us to express a multi-trace unitary matrix model in terms of a single trace matrix model \cite{Alvarez-Gaume:2006jg}.

Since the effective action (\ref{genaction}) is a polynomial in $\tr\, U^{i}$, $\tr\, U^{\dagger i}$, we can use the standard Gaussian trick twice to write the partition function (\ref{genz}) as
\bea
\label{fullint}
Z=\biggl( {N^4\over 2\pi^2}\biggr)^{p} \int   \prod_{i=1}^p dg_i \, d\bar g_i  \, d\mu_i \, d\bar \mu_i \exp (N^2 S_{\rm eff})
\eea
where
\bea
\label{EffAction}
S_{\rm eff}&=&-\sum_{j=1}^p a_j \mu_j \bar \mu_j +{\rm i}\sum_j (\mu_j \bar g_j + \bar \mu_j g_j) + \\
\nn & & \sum_{\vec k, \vec k'}  \alpha_{\vec k , \vec k'} (-{\rm i})^{|\vec k| + |\vec k'|} \Upsilon_{\vec k} (\bar \mu) \Upsilon_{\vec k'}(\mu) + F(g_k , \bar g_k ).
\eea
In the above formula we have introduced the definition
\be
\Upsilon_{\vec k}(\mu)=\prod_j \mu_j^{k_j}.
\ee
and the free energy $F(g_k, \bar g_k)$ is defined by
\bea
\label{standardmodel}
\exp (N^2 F(g_k , \bar g_k)) &=& \int [d\, U] \exp\biggl\{ N \sum_{i\ge 1 } ( g_i  {\tr}\, U^i +  \bar g_i {\tr}\, U^{\dagger i} ) \biggr\},
\eea
\par

\section{Critical behavior in the matrix model}
 The eigenvalues of an unitary matrix $U$ are the complex numbers $e^{i\theta_i }$.\footnote{ Phase structure of a generic unitary matrix model has been discussed in \cite{Mandal:1989ry}}. In the large $N$ limit, we can consider an eigenvalue density $\rm \rho(\theta)$ defined on the unit circle by,
 
\bea
\rho(\theta) &=& \frac{1}{N}\sum_{i=1}^{N}\delta (\theta -\theta_i)= \sum_{n}\exp(i n\theta)\frac{1}{N} \tr \, U^n
\eea
The density function is non-negative and normalized,
\bea
\int \rm \rho(\theta) d\theta =1 \\
\rm \rho(\theta) \ge 0
\eea
It is well known that in the limit of $N\rightarrow \infty$, $\rho (\theta )$ can develop gaps, i.e. it can be non-zero only in bounded intervals. For example, in the case of a single gap when $\rho (\theta )$ is non-zero only in the interval $(-{\theta_0 \over 2},\frac{\theta_{0}}{2})$, it is given by the classical formula 

\bea
\rho(\theta)= f(\theta)\sqrt{\sin^2 {\theta_0 \over 2} -\sin^2{ \theta \over 2}}
\label{gapdistri}
\eea

A well known example of a $\rho(\theta)$ which does not have a gap is

\bea
\rho(\theta)= \frac{1}{2\pi}(1+a\cos(\theta)), \quad a<1
\label{ungapdistri}
\eea

At $a=1$, $\rho(\pi)=0$, and a gap will begin to open. 
For $a>1$ the functional form of $\rho(\theta)$ is as given by (\ref{gapdistri}). In general the condition $\rho(\pi)=0$ defines a critical surface in the space of couplings of the effective action. 
The saddle point distribution of the eigenvalues of the matrix $U$ may or may not have a gap, depending on the values of parameters $g_k$ in (\ref{standardmodel}). The opening/closing of the gap in the eigenvalue distribution signals the Gross-Witten-Wadia (GWW) third order phase transition in the matrix model \cite{Gross:1980he}, \cite{Wadia:1979vk}, \cite{Wadia:1980cp}.

In the large $N$ expansion, the functional dependence of $F(g_k ,\bar g_k )$ on $g_k,\bar g_k$  depends on the phase, and we quote from the known results \cite{Goldschmidt:1979hq}, \cite{Periwal:1990gf}, \cite{Periwal:1990qb},
\bea
\label{Nexpansion}
N^2 F(g_k,\bar g_k) &=& N^2 \sum_k \frac{kg_k \bar g_k}{4} + e^{-2Nf(g_k,\bar g_k)}\sum_{n=1}^{n=\infty}\frac{1}{N^n}F_{n}^{(1)}, \quad {\rm ungapped} \\
\nn N^2 F(g_k,\bar g_k) &=& N^2 \sum_k \frac{kg_k \bar g_k}{4}+\sum_{n=0}^{n=\infty} N^{-\frac{2}{3}n}F_{n}^{(2)}, \quad g-g_c\sim o( N^{-\frac{2}{3}}) \\
\nn  N^2 F(g_k,\bar g_k) &=& N^2 G(g_k,\bar g_k)+\sum_{n=1}^{n=\infty}\frac{G^{(n)}}{N^2},\quad {\rm gapped}
\eea
$f(g_k,\bar g_k),F_{n}^{(1)},F_{n}^{(2)}$ and $G^{n}(g_k,\bar g_k)$ are calculable functions using standard techniques of orthogonal polynomials. 

Two comments are in order:

i) In the above, we have assumed for simplicity that the eigenvalue distribution has only one gap. (In principle we cannot exclude the possibility of a multi gap solution. But here, since we are interested in the critical phenomena that results when the gap opens (or closes) we will concentrate on the single gap solution.) Near the boundary of phases, the functions $F_{n}(g)$ and $G_{n}(g)$ diverge. It is well known that in the leading order $N$, $F(g_k,\bar g_k)$ has a third order discontinuity at the phase boundary. This non-analytic behavior is responsible for the large $N$ GWW type transition. In the $o(N^{-\frac{2}{3}})$ scaling region near the phase boundary (the middle expansion in (\ref{Nexpansion})) this non-analytic behavior can be smoothened by the method of double scaling. This smoothening is important for our calculation of the double scaled partition function near the critical surface. 
\par

\par
2) In the gapped phase of the matrix model, $F(g_k,\bar g_k)$ has a standard expansion in integer powers of  $\frac{1}{N^2}$, which becomes divergent as one approaches the critical surface. In the double scaling region (\ref{Nexpansion}) $(g-g_c) \sim \CO(N^ {- {2\over 3}})$, and the the perturbation series (\ref{Nexpansion}) is organized in an expansion in powers of $N^{-\frac{2}{3}}$. The reason for the origin of such an expansion is not clear from the viewpoint of the bulk string theory. However, it is indeed possible to organize the perturbation series, in the scaling region, in terms of integral powers of a renormalized coupling constant. We will come back to this point later. In the ungapped phase the occurrence of $o(e^{-N})$ terms is also interesting. Here too we lack a clear bulk understanding of the non-perturbative terms which naturally remind us of the D-branes.

\subsection{Critical surface of the large N phase transition}

We now describe the critical surface in the space of couplings across which there is a GWW phase transition.
\par
From (\ref{Nexpansion}) we can easily find the density of eigenvalues in the ungapped phase. 
\bea
\rho(\theta)&=&\frac{1}{2\pi}(1+\sum_{k}(k g_k\exp(ik\theta)+ k \bar g_k\exp(ik\theta)) \\
\nn \rm and \quad \rho_k&=&k g_k
\label{rhoungap}
\eea 
For a set of real $g_k$, the lagrangian (\ref{standardmodel}) is invariant under $U \rightarrow U^{\dagger}$. We will assume that the gap opens at $\theta=\pi$ according to $\rho(\pi - \theta)\sim (\pi - \theta)^2$, which characterizes the first critical point\footnote{In general the mth critical point is characterized by $\rho(\pi - \theta)\sim (\pi - \theta)^{2m}$.}. At the boundary of the gapped-ungapped phase (critical surface) we have $\rho(\pi)=0$. In terms of the critical fourier components $\rho^{c}_k$, it is the equation of a plane with normal vector $\tilde D_k = (-1)^k$
\bea
\sum_{k=-\infty}^{\infty} (-1)^k (\rho^{c}_k+\bar \rho^{c}_k)= -1 
\label{critsur1}
\eea
Now since $\rho^{c}_k = kg^{c}_k$ (up to non-perturbative corrections),
we get the equation of a plane 
\bea
\sum_{k=-\infty}^{\infty} (-1)^k k (g^{c}_k+\bar g^{c}_k)= -1 
\label{critsur}
\eea
where $g^{c}_k$ are the values of $g_k$ at the critical plane. Since the metric induced in the space of $g_k$ from the space of $\rho_k$ is $G_{k,k'} = k^2\delta_{k,k'}$, the vector that defines this plane is 
\be
C_k = \frac {(-1)^k}{k}
\ee
We mention that the exact values of $g^c_k$ where the thermal history of the small blackhole intersects the critical surface are not known to us as we do not know the coefficients of the effective lagrangian. However this information, which depends on the details of dynamics, does not influence the critical behavior. The information where the small blackhole crosses the critical surface is given by the saddle point equations, which are in turn determined by the $o(N^2)$ part of the action (\ref{EffAction}).
\par

\section{Saddle point equations at large N}
\par
The saddle points of (\ref{EffAction}) corresponding to the N=4 SYM theory are in correspondence with the bulk supergravity (more precisely IIB string theory) saddle points. For example, the $AdS_5 \times S^5$ geometry corresponds to a saddle point such that $\langle \tr \, U^n\rangle=0$ $\forall n \neq 0$. Hence the eigenvalue density function is a uniform function on the circle. Now, depending on the co-efficients in (\ref{EffAction}) the saddle point $\vev{\tr \, U^n}$ can have a non-uniform gaped or ungapped eigenvalue density profile. Changing the values of the coefficients, by varying the temperature, may open or close the gap and lead to non-analytic behavior in the temperature dependence of the free energy at $N=\infty$  We will interpret this phenomenon, the GWW transition, as the string-blackhole transition. As we shall see this non-analytic behavior can be smoothened out by a double scaling technique in the vicinity of the phase transition.

The $o(N^2)$ formulas for the free energy leads to the large $N$ saddle point equations for the multi-trace matrix model (\ref{fullint}). \par
\begin{figure}
\includegraphics[height=6cm]{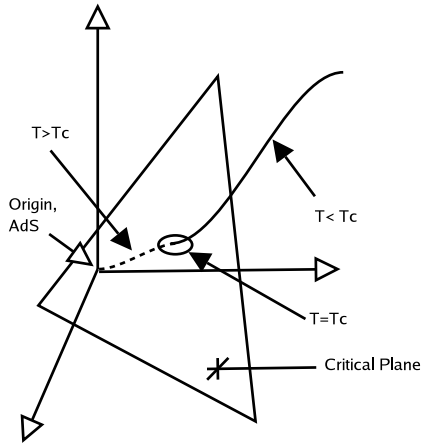}
\label{criticalsurface}
\caption{Critical plane in the $\rho$ space and thermal history of the saddle point}
\end{figure}

By the AdS/CFT correspondence the solutions to the saddle point equations are dual to supergravity/string theory solutions, like $AdS_5 \times S^5$ and various $AdS_5 \times S^5$ blackholes. The number and types of saddle points and their  thermal histories depends on the dynamics of the gauge theory (i.e. on the numerical values of the parameter $a_j$ and $ \alpha_{\vec k, \vec k'}$, which in turn are complicated functions of $\lambda$ and $\beta$). These issues have been discussed in the frame work of simpler models in \cite{Alvarez-Gaume:2005fv}, where the first order confinement/deconfinement transition and its relation with the Hawking-Page type transition in the bulk has also been discussed. Here we will not address these issues, but focus on the phenomenon when an {\it unstable saddle point} crosses the critical surface (\ref{critsur}). (see Fig(\ref{criticalsurface}).)

\par
In a later section we will use the AdS/CFT correspondence to argue that in the strongly coupled gauge theory, a 10 -dimensional ``small blackhole'' saddle point reaches the critical surface $\rho(\pi)=0$. The interpretation of this phenomenon in the bulk string theory, as a blackhole to excited string transition will also be discussed. 

\section{Double scaled partition function at crossover}\label{finalsection}

We will assume that the matrix model (\ref{standardmodel}) has a saddle point which makes a gapped to ungapped transition as we change the parameters of the theory($\alpha_{\vec k, \vec k'}^c$,$a_j$) by tuning the temperature $\beta^{-1}$. We will also assume that, this saddle point has one unstable direction which corresponds to opening the gap as we lower the temperature. These assumptions are motivated by the fact that the small (euclidean) Schwarzchild blackhole crosses the critical surface and merges with the $AdS_5 \times S^5$ and that it is an unstable saddle point of the bulk theory. To calculate the doubled scaled partition function near this transition point, we basically follow the method used in \cite{Alvarez-Gaume:2005fv}. We expand the effective action (\ref{standardmodel}) around the $1{\it st}$ critical point, and we simultaneously expand the original couplings $a_j$, $ g_j$, $\bar g_j$ and $\alpha_{\vec k, \vec k'}$ around their critical values  $a_j^c$, $\beta_j^c$, $g_j^c=0$, and $\alpha_{\vec k, \vec k'}^c$. For clarity we define  
\bea
P(\mu, \bar \mu, \alpha)=\sum_{\vec k, \vec k'} \alpha_{\vec k , \vec k'} (-{\rm i})^{|\vec k| + |\vec k'|} \Upsilon_{\vec k} (\bar \mu) \Upsilon_{\vec k'}(\mu)
\eea

 We also introduce the column vectors,

\be
\label{varias}
 \mu =  \left( \begin{matrix}  \mu_j \\
                                  \bar\mu_j \end{matrix} \right), \quad
 A= \left( \begin{matrix} a_j \\
                                  \alpha_{\vec k, \vec k'}   \end{matrix} \right), 
                             \quad
                             g= \left( \begin{matrix} g_j \\
                                  \bar g_j \end{matrix} \right)                                   
\ee
and expand the above mentioned vector variables
\bea
 g- g^c &=& N^{-\frac{2}{3}}\tilde{t} \\
\nn \mu- \mu^c &=& N^{-{4 \over 3}} n \\
\nn  A - A^{c} &=& \tilde{g}  N^{-\frac{2}{3}} \alpha
\label{scaling}
\eea
where $\tilde{g}=N^{\frac{2}{3}}(\beta-\beta_c)$ and $\alpha={\partial A \over \partial \beta}|_{\beta=\beta_c}$.  The expansion of the co-efficients $a_j$ and $\alpha_{\vec k, \vec k'}^c$
 are proportional to the deviation of the tuning parameter $\beta$ from its critical value, i.e. $\tilde{g}=N^{\frac{2}{3}}(\beta_c-\beta)$.

Putting the above expression in ({\ref{expg}) and performing the gaussian integral we get the final result,
\bea
Z \sim  i({\rm det}(H))^{-{1\over 2}}\exp F(C \cdot \tilde{t})
\label{finalresult}
\eea
We have assumed that the Hessian $H$ does not have a zero mode, but the one negative eigenvalue accounts for the $i$ in front of (\ref{finalresult}). We have no independent derivation of the existence of the negative eigenvalue except that the dual blackhole has exactly one unstable direction in the euclidean signature.

\par
Note that $C\cdot \tilde{t} = t$ is a parameter along the vector $C$ which is normal to the critical surface. It can be proved from the discrete recursion relations of the matrix model that the function $F(t)$ in (\ref{finalresult}) is given in terms of the Painleve II function $f(t)$,
\bea 
\ddel{F}{t} =-f^2(t) 
\eea
where $f(t)$  satisfies the Painleve II equation,
\bea
\frac{1}{2}\ddel{f}{t}=tf+f^3
\eea
The exact form of $F(t)$ is not known but it is known that it is a smooth function with the following asymptotic expansion. 
\bea
 F(t) &=&  {t^3 \ov 6}
 - {1 \ov 8} \log (-t) - {3 \ov 128 t^3}
 + {63 \ov 1024 t^6} + \cdots  ,\quad -t \gg 1 \\
\nn F(t) &=& {1 \ov 2 \pi} e^{-  {4 \sqrt{2} \ov 3} t^{3 \ov 2}}
 ( - {1 \ov 8 \sqrt{2} t^{3 \ov 2} } + {35 \ov 384 t^3} -
 {3745 \ov 18432 \sqrt{2} t^{9 \ov 2}} + \cdots ), \quad  t \gg 1 
 \eea

\par
The $o(1)$ part of the partition function, (\ref{finalresult}) is universal in the sense that the appearance of the function $F(t)$, does not depend on the exact values of the parameters of the theory. Exact values of the couplings and the $o(N^2)$ part of the partition function determine where the thermal history crosses the critical surface (\ref{critsur}). However the form of the function $F$ and the double scaling limit of (\ref{scaling}) are independent of the exact values of $g^{c}_k$. They only depend on the fact that one is moving away perpendicular to the critical surface. This is the reason why in \cite{Alvarez-Gaume:2005fv} we obtained exactly the same equation when $g^{c}_1\neq 0$ but all other $g^{c}_k=0$. 
\subsection{Condensation of winding modes at the crossover}

We will now discuss the condensation of the winding Polyakov lines in the crossover region. In the leading order in large N it is not difficult to see that $\rho_k^c = kg_k^c$, where $\rho_k = <\frac{1}{N}\tr \, U^{k}>$. In order to calculate subleading corrections it can be easily seen that all the $\rho_k$'s condenses in the scaling region, 
\bea
\vev{N^{\frac{2}{3}}(\rho_k-\rho^{ug}_k)} = C_k \frac{dF}{dt}
\label{condensation}
\eea
where $\rho^{ug}_k=kg_k$.  
This smoothness of the expectation value of the $\rho_k$'s follows from the smooth nature of $F(t)$. 
The derivative of $F(t)$ diverges as $t \rightarrow -\infty$ and goes to zero as $t \rightarrow \infty$. This behavior tallies with the condensation of winding mode in one phase (the gapped phase) and the non-condensation of winding modes in the ungapped phase. The condensation of the winding modes also indicates that the $U(1)$ symmetry (which is the $Z_N$ symmetry of the $SU(N)$ gauge theory in the large N limit) is broken at the crossover, but restored in the limit $t \rightarrow \infty$.

\section{Applications to the small blackhole-string transition}
We now apply what we have learned about the matrix model (gauge theory) GWW transition and its smoothening in the critical region to the blackhole-string transition in the bulk theory. The first step is to identify the matrix model phase in which the blackhole or for that matter the supergravity saddle points occur. We will argue that they belong to the gapped phase of the matrix model. This inference is related to the way perturbation theory in $\frac{1}{N}$ is organized in the gapped, and ungapped phase as discussed in (\ref{Nexpansion}). Note that it is only in the gapped phase, that the $\frac{1}{N}$ expansion is organized in powers of $\frac{1}{N^2}$, exactly in the way perturbation theory is organized around classical supergravity solutions in closed string theory. Hence at the strong gauge theory coupling($\lambda \gg 1$), it is natural to identify the small 10 dimensional blackhole with a saddle point of the equations of motion obtained by using $F(g_k,\bar g_k)$  corresponding to the gapped phase. \footnote{A saddle point of the weakly coupled gauge theory may also exist in the gapped phase. With a change in the temperature the saddle point can transit through the critical surface. Using the results of \cite{Alvarez-Gaume:2005fv}, it is easy to see that this is precisely what happens for the perturbative gauge theory discussed in \cite{Aharony:2005bq}. We note that in the corresponding bulk picture since $l_s >> R_{AdS}$, the supergravity approximation is not valid. It would be interesting to understand the bulk interpretation in this case.} One can associate a temperature with this saddle point which would satisfy $l_s^{-1} \gg T \gg R^{-1}$. 

As the temperature increases towards  $l_s^{-1}$, one traces out a curve (thermal history) in the space of the parameters $a_i, \alpha _{k, k'}$ of the effective theory. 
One can also say that a thermal history is traced in the space of $\rho _i =\langle \frac{1}{N}\tr \, U^i\rangle$, which depends 
on the parameters of the effective theory. We will now make the reasonable assumption that the thermal history, at a temperature $T_c \sim l_s^{-1} $, 
intersects the critical surface (\ref {critsur1}) (equivalently the plane (\ref {critsur}) and then as the temperature increases further it reaches the point 
$\rho _i =\langle \frac{1}{N}\tr \, U^i \rangle = 0$, which corresponds to $AdS_5\times S^5$. Once the thermal history crosses the critical surface, the gauge theory 
saddle points are controlled by the free energy of the ungapped phase in (\ref{Nexpansion}). The saddle points of eqns. (\ref{saddleqs1}) which were obtained 
using this free energy do not correspond to supergravity backgrounds, because the temperature, on crossing the critical surface is very high $T \gtrsim l_s^{-1}$. Besides 
this the free energy in the gapped phase has unconventional exponential factors (except at $g_k=0$ which corresponds to $AdS_5\times S^5$). It is likely that these 
saddle points define in the correspondence, exact conformal field theories/non-critical string theories in the bulk. Neglecting the exponential corrections $exp(-N)$, it seems 
reasonable, by inspecting the saddle point equations, that in this phase the spectrum would be qualitatively similar to that around $\rho _i = 0$. Since this corresponds to 
$AdS_5\times S^5$, we expect the fluctuations to resemble a string spectrum.

As we saw in the previous section, our techniques are good enough only to compute a universal $o(1)$ part of the partition function in the vicinity of the critical surface. The exact solution of the free energy in the transition region in (\ref{Nexpansion}) enabled us to define a double scaling limit  
in which the non-analyticity of the partition function could be smoothened out, by a redefinition of the string coupling constant according to 
$\tilde{g}=N^{\frac{2}{3}}(\beta_c-\beta)$. This smooth crossover corresponds to the blackhole crossing over to a state of strings corresponding to the ungapped phase. 

We have also computed the vev of the scaling operator and hence at the crossover the winding modes $\rho _i = \langle \frac{1}{N}\tr \, U^i\rangle$ condense (\ref{condensation}). They also 
have a smooth parametric dependence across the transition. This phenomenon in the bulk theory may have the interpretation of smooth topology change of a 
blackhole spacetime to a spacetime without any blackhole and only with a gas of excited string states. However in the crossover region a geometric spacetime 
interpretation is unlikely. We may be dealing with the exact description of a non-critical string in 5-dims. in which only the zero mode along the $S^3$ directions 
is taken into account. This interpretation is inspired by the fact that the free energy $F(t)$ also describes the non-critical type 0B theory as was already discussed in \cite{Alvarez-Gaume:2005fv}. 
\par
\section{Density of states and the singularity of the Lorentzian blackhole}

The resolution of the singularity of a blackhole, that occurs in its general relativity description, is a fundamental question in string theory. We interpret our result in favor this resolution in the gauge theory. 
Since the partition function, in an appropriate scaling limit, is a smooth function of the renormalized coupling constant $\tilde g$, at the crossover between the gapped and ungapped phase, it is clear that the density of states $\rho(E)$ also inherits the same property. Since $\rho(E)$ is as well a quantity that has meaning when the signature of time is Lorentzian, it would imply that the blackhole-string crossover in the Lorentzian signature is also smooth. This is an interesting conclusion especially because we do not know the AdS/CFT correspondence for the small Lorentzian blackhole. The Lorentzian section of the blackhole has a singularity behind the horizon. Since the gauge theory should also describe this configuration, a smooth density of states in the crossover may imply that the blackhole singularity is resolved in the gauge theory. A more direct Lorentzian calculation is will strengthen this conclusion.


\section{Acknowledgment}
I would like to acknowledge and thank the organizers of the 
12th Regional conference in Islamabad, especially Prof. Faheem Hussein, for efforts that made the conference and the visit to Pakistan a memorable and wonderful experience. This research is supported in part by the J. C. Bose Fellowship of the Dept. of Science and Technology, Govt. of India.

\par

\end{document}